\documentclass[iop]{emulateapj}

\shorttitle{Cyanogen in NGC 1851 Giants}
\shortauthors{Campbell et al.}

\begin{document}


\title{Cyanogen in NGC 1851 RGB and AGB Stars: Quadrimodal Distributions}

%
\author{S. W. Campbell\altaffilmark{1}}
\affil{Monash Centre for Astrophysics, PO Box 28M, Victoria 3800, Australia}
\email{simon.campbell@monash.edu}
\author{D. Yong\altaffilmark{2} and E. C. Wylie-de Boer\altaffilmark{2}}
\affil{Research School of Astronomy and Astrophysics, 
            Australian National University, 
            Weston, ACT 2611, Australia}
\email{david.yong@anu.edu.au}
\author{R. J. Stancliffe\altaffilmark{1,2,3}}
\affil{Argelander-Insitut f\"{u}r Astronomie, Universit\"{a}t Bonn, Auf dem
  H\"{u}gel 71, 53121 Bonn, Germany}

\author{J. C. Lattanzio\altaffilmark{1} and G. C. Angelou\altaffilmark{1}}
\author{V. D'Orazi\altaffilmark{4,1}}
\affil{Department of Physics \& Astronomy, Macquarie University, Balaclava Rd, North Ryde, Sydney, NSW 2109, Australia}
\author{S. L. Martell\altaffilmark{5}}
\affil{Australian Astronomical Observatory, North Ryde, NSW 2113, Australia}
\author{F. Grundahl\altaffilmark{6}}
\affil{Department of Physics and Astronomy, 
           Aarhus University, Ny Munkegade, 8000 Aarhus C, Denmark}
\and
\author{C. Sneden\altaffilmark{7}}
\affil{Department of Astronomy and McDonald Observatory, 
           University of Texas, Austin, TX 78712, USA}
%



\begin{abstract}
The Galactic globular cluster NGC 1851 has raised much interest since HST
photometry revealed that it hosts a double subgiant branch. Here we report
on our homogeneous study into the cyanogen (CN) bandstrengths in the RGB
population (17 stars) and AGB population (21 stars) using AAOmega/2dF
spectra with R $\sim 3000$. We discover that NGC 1851 hosts a
\emph{quadrimodal} distribution of CN bandstrengths in its RGB \emph{and}
AGB populations.  This result supports the merger formation scenario
proposed for this cluster, such that the CN quadrimodality could be
explained by the superposition of two `normal' bimodal populations. A small
sample overlap with an abundance catalogue allowed us to tentatively
explore the relationship between our CN populations and a range of
elemental abundances. We found a striking correlation between CN and
[O/Na]. We also found that the four CN peaks may be paired -- the two
CN-weaker populations being associated with low Ba and the two CN-stronger
populations with high Ba. If true then s-process abundances would be a good
diagnostic for disentangling the two original clusters in the merger
scenario. More observations are needed to confirm the quadrimodality, and
also the relationship between the subpopulations. We also report CN results
for NGC 288 as a comparison. Our relatively large samples of AGB stars show
that both clusters have a bias towards CN-weak AGB populations.
\end{abstract}

\keywords{Stars: AGB and post-AGB --- globular clusters: general ---
  globular clusters: individual(NGC 1851, NGC 288)}

\section{Introduction}\label{sec:intro}

Galactic globular clusters (GCs) are no longer thought to be perfectly
homogeneous, simple stellar populations. Although almost all are chemically
homogeneous with respect to Fe and heavier elements (omega Cen, M22, Terzan
5 and NGC 1851 being exceptions), it has long been known that GCs show
large star-to-star abundance variations for light elements (e.g., C, N, O,
Na, Mg, Al; see reviews by
\citealt{Kraft94,Gratton04,Gratton12multi}). These inhomogeneities are
considered anomalous because they are seen in very few halo field stars of
similar metallicity (\citealt{GSC00,MG10}). Studies of all phases of
evolution, including the red giant branch (RGB), main sequence (MS) and
subgiant branch (SGB; e.g. in NGC 6752, \citealt{GBB01}), have shown the
same anomalies. This suggests that many of the abundance variations arose
in the early phases of cluster evolution.

Recently it was discovered that the globular cluster NGC 1851 has a double
SGB, whereby two evolutionary sequences are clearly visible (HST
photometry, \citealt{Milone08,Milone09}). The RGB has also been shown to
split into two when using particular filters \citep{Han09}. Spectroscopic
observations of this cluster show bimodality in s-process abundances
\citep{Yong08,Villanova10,Gratton12SGB} and a small spread in [Fe/H] (rms
scatter $\sim 0.05$ dex, \citealt{Carretta11}).  \citealt{Gratton12SGB}
report that the two SGB populations have slightly different heavy element
contents on average.  \cite{Carretta11} find they can split their RGB
sample into a metal-rich and metal-poor population based on the Fe-Ba
plane, and that each population has its own O-Na anticorrelation. This ties
in well with the horizontal branch (HB) observations of \cite{Gratton12HB}
who also report two separate O-Na anticorrelations. As an explanation for
these abundance anomalies, as well as the bimodal HB, it has been suggested
that NGC 1851 may be a product of a merger between two GCs
\citep{Vandenbergh96,Catelan97,Carretta10}. \cite{Bekki12} recently showed
that a merger scenario for NGC 1851 is dynamically plausible. For reviews
on the phenomenon of multiple populations in GCs see
e.g. \cite{Piotto09}, \cite{Martell11} and \cite{Gratton12multi}.

One of the first inhomogeneities discovered in globular clusters was that
of the molecule cyanogen (CN, often used as a proxy for nitrogen). A
picture of `CN-bimodality' emerged in the 1970s and 80s
\citep{Hesser78,Norris79,Cottrell81} whereby stars in one population show
weak absorption by CN (`CN-weak' stars) and stars in the the other show
strong absorption by CN (`CN-strong' stars). This has been observed in
most, if not all, clusters. With the recent interest in NGC 1851 there have
been a couple of studies of CN, on the MS \citep{Pacino10} and the two SGBs
\citep{Lardo12}. There does however appear to be a dearth of studies of CN
in giants in NGC 1851 -- here we report on observations focusing on
cyanogen band strengths in the RGB and AGB stars of NGC 1851.

\section{Stellar Sample, Observations and Data Reduction}\label{sec:obs}

\begin{figure}[!ht]
\centering
  \includegraphics[width=0.8\columnwidth]{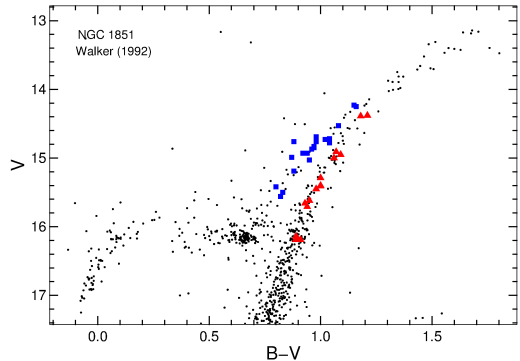}
  \caption{{\footnotesize The stellar sample. Small dots are all the stars
      from the \cite{Walker92} CMD. Filled triangles (red) are our sample
      of 17 RGB stars, filled squares (blue) are our 21 AGB stars.}}
  \label{fig:cmdselect}
\end{figure}

Our stellar sample was taken from the BV photometry catalogue of
\cite{Walker92}. This catalogue was chosen because the photometry is
precise enough to distinguish between the RGB and AGB populations and
because it provides accurate astrometry, an important feature for
multi-fibre spectroscopy. The chosen sample of stars was cross-matched with
the 2MASS catalogue \citep{SCS06}, thus all of our sample stars are
actually 2MASS objects, with positions accurate to $\sim 0.2$ arcsec. Since
the two giant branches merge in the CMD at higher luminosities we limited
our RGB and AGB samples to V $>14.2$.  We show our program stars against
the \cite{Walker92} CMD in Figure \ref{fig:cmdselect} and provide a list in
Table \ref{tab:observations}.

Our observational data were taken over the second halves of the nights of
the 5th, 7th, 8th and 9th of September 2009 at the AAT using the
multi-fibre spectrograph, AAOmega/2dF (\citealt{LCT02,SBG04,SSS06}). A
total of 9 hours of exposures were taken, using 3 field plate setups. The
1700B grating was used on the blue arm of the spectrograph, which gave a
spectral coverage of $3755-4437\textrm{\AA}$ and includes the violet CN
bands around $3850-3880$\AA. Spectral resolution in this region was R $\sim
3000$.

Data reduction was carried out using the 2dF pipeline software $2dFdr$
(version 3.211, April 2009) provided by the AAO.  Tram-map fits to the
multiple spectra from each plate were checked by eye, as were the arc
reductions and final reduced science spectra. Our final sample of spectra
contained 17 RGB and 21 AGB stars.

To quantify the CN band strengths in each spectrum we used the S(3839) CN
index of \cite{NCF81} which compares a section of the CN bands with a
neighboring pseudo-continuum (Eqn. \ref{eqn1}). IRAF was used to measure
the integrated fluxes of Equation \ref{eqn1} for all the program stars.
\begin{equation}
\label{eqn1}
S(3839) = -2.5 \log \frac{\int_{3846}^{3883} \! I_\lambda \, d\lambda}{\int_{3883}^{3916} \! I_\lambda \, d\lambda}
\end{equation}

\section{Results and Discussion}\label{sec:results}

\begin{figure}
\centering
\includegraphics[width=0.8\columnwidth]{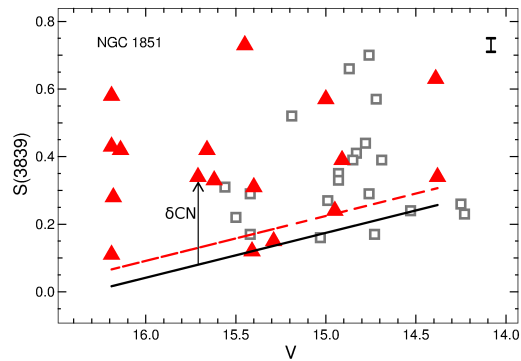}
\caption{{\footnotesize Measured S(3839) CN index versus magnitude for the
    NGC 1851 stars. Filled triangles (red) are RGB stars, open squares
    (grey) are AGB stars. The dashed line shows a least-squares fit to the
    5 RGB stars at the lower envelope of the distribution. The solid line
    is the same but offset so that the zero level of $\delta$S(3839)
    ($\delta$CN) is coincident with the star with the lowest
    $\delta$CN. The definition of $\delta$CN is shown by the arrow. A
    characteristic error bar for S(3839) is shown at top right (see text
    for details).}}
\label{fig:rawdata}
\end{figure}

\begin{figure*}[!t]
\centering
\includegraphics[width=\textwidth]{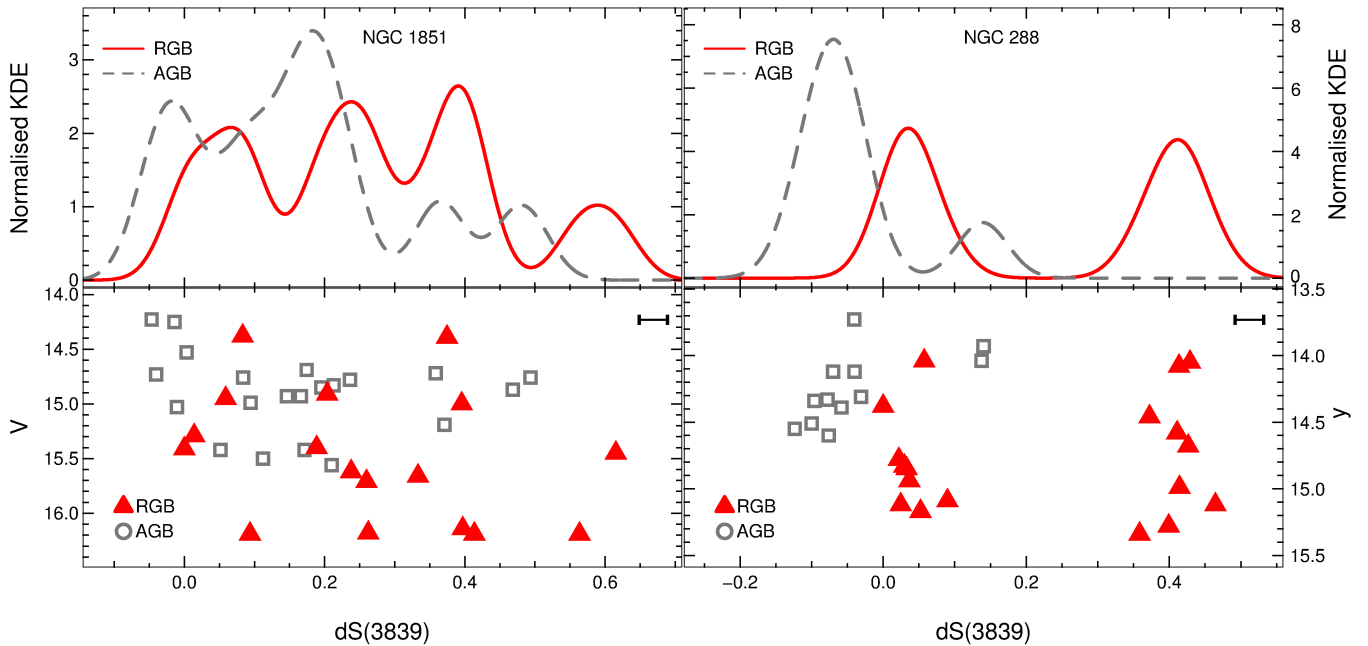}
\caption{\footnotesize{\textit{Left-hand Bottom Panel:} $\delta$S(3839) CN
    index versus V magnitude for NGC 1851 (see Fig. \ref{fig:rawdata} for
    the definition of $\delta$S(3839)). \textit{Left-hand Upper Panel:}
    Normalised kernel density estimate (KDE) histogram (Gaussian kernel,
    bandwidth $=0.035$) of $\delta$S(3839). A clear CN quadrimodality can
    be seen on the RGB and the AGB.  A characteristic error bar for S(3839)
    is shown at top right of the lower panel. \textit{Right-hand Panels:}
    Same as left-hand panels but for NGC 288, for comparison. The bandwidth
    for the Gaussian kernel is the same as that used for the NGC 1851
    data. In this case photometry ($uvby$) for the stellar sample selection
    was provided by F. Grundahl (private communication; \citealt{GCL99})} }
\label{fig:doubleplot}
\end{figure*}

In Table \ref{tab:observations} we list the S(3839) measurements for each
star, while in Figure \ref{fig:rawdata} we plot them versus V magnitude. CN
absorption is known to have a temperature dependence so we have
`de-trended' the data in the same manner as in previous CN studies
(e.g. \citealt{NCF81,Ivans99,Martell08}) by fitting a line to the lower
envelope of the observations. The value $\delta$S(3839) (hereafter
$\delta$CN) is then the distance from this line to each data point. The
resultant $\delta$CN values are shown in the lower panel of Figure
\ref{fig:doubleplot} and listed in Table \ref{tab:observations}. Errors in
wavelength calibration or doppler offsets due to velocity dispersion were
checked and found to be of order $\sim 10^{-3}$ in $\delta$CN. The much
larger characteristic error bar for S(3839) given in the figures ($\pm
0.02$) reflects the typical differences between measurements of S(3839) in
two observations of the same star. These pairs of observations were taken
on different nights and with different field plates (other clusters in our
broader observational campaign were used for this, but the data were taken
during the same timeframe as for NGC 1851). We found that this was by far
the largest source of error. This is probably to be expected since it
reflects the combination of many sources of error, including the
uncertainties of fibre placement, fibre throughput, slight pointing errors,
seeing variability, as well as errors in the data reduction (for example).

\begin{table*}
\caption{\footnotesize{List of target stars for NGC 1851. IDs (column 2), V magnitudes and
  B-V values are from \cite{Walker92}. Column 4 IDs are from \cite{Carretta11}
  and show the overlap between studies. In column 7 are our raw CN band
  strength measurements, S(3839), and in column 8 our de-trended CN index
  values $\delta$S(3839).}}
\label{tab:observations}
\begin{center}
\footnotesize{
    \begin{tabular}{clcccccc}
        \tableline\tableline
TYPE 	&	 ID(Walk.) 	&  ID(2MASS) 	&
ID(Carr.)	&	 V 	&	 B-V 	&	 S(3839) 	&
$\delta$S(3839)	\\ \tableline
RGB 	&	24	&	 05134832-4003151 	&	--	&	15.40	&	1.00	&	0.31	&	0.19	\\
RGB 	&	28	&	 05134897-4001199 	&	41113	&	15.71	&	0.94	&	0.34	&	0.26	\\
RGB 	&	71	&	 05135414-4003038 	&	32112	&	15.45	&	0.98	&	0.73	&	0.62	\\
RGB 	&	79	&	 05135462-4005094 	&	--	&	15.29	&	1.00	&	0.15	&	0.01	\\
RGB 	&	120	&	 05135671-4001016 	&	--	&	15.00	&	1.06	&	0.57	&	0.40	\\
RGB 	&	151	&	 05135828-3959586 	&	--	&	16.19	&	0.90	&	0.43	&	0.41	\\
RGB 	&	160	&	 05135866-4000178 	&	--	&	16.14	&	0.89	&	0.42	&	0.40	\\
RGB 	&	161	&	 05135867-4000120 	&	44803	&	14.91	&	1.07	&	0.39	&	0.20	\\
RGB 	&	162	&	 05135862-3959242 	&	46228	&	16.19	&	0.91	&	0.11	&	0.09	\\
RGB 	&	208	&	 05135977-4001374 	&	--	&	15.66	&	0.93	&	0.42	&	0.33	\\
RGB 	&	368	&	 05140259-4000220 	&	44414	&	15.62	&	0.95	&	0.33	&	0.24	\\
RGB 	&	441	&	 05140365-4001596 	&	--	&	16.18	&	0.89	&	0.28	&	0.26	\\
RGB 	&	1028	&	 05141052-3958095 	&	47385	&	15.41	&	1.00	&	0.12	&	0.00	\\
RGB 	&	1256	&	 05141724-4002080 	&	37070	&	14.38	&	1.21	&	0.34	&	0.08	\\
RGB 	&	1284	&	 05141956-4004055 	&	26801	&	14.39	&	1.18	&	0.63	&	0.37	\\
RGB 	&	1286	&	 05141947-4000076 	&	44939	&	14.95	&	1.09	&	0.24	&	0.06	\\
RGB 	&	1323	&	 05142281-4001551 	&	38215	&	16.19	&	0.89	&	0.58	&	0.56	\\
AGB 	&	182	&	 05135918-4002496 	&	--	&	14.93	&	0.92	&	0.35	&	0.17	\\
AGB 	&	222	&	 05140019-4002291 	&	--	&	14.69	&	0.98	&	0.39	&	0.17	\\
AGB 	&	245	&	 05140068-4003239 	&	30315	&	14.53	&	1.08	&	0.24	&	0.00	\\
AGB 	&	430	&	 05140355-4002499 	&	--	&	15.03	&	0.95	&	0.16	&	-0.01	\\
AGB 	&	506	&	 05140446-4003113 	&	--	&	14.76	&	0.88	&	0.29	&	0.08	\\
AGB 	&	572	&	 05140508-4002278 	&	--	&	15.56	&	0.82	&	0.31	&	0.21	\\
AGB 	&	633	&	 05140584-4002126 	&	--	&	14.25	&	1.16	&	0.26	&	-0.01	\\
AGB 	&	680	&	 05140659-4002026 	&	--	&	14.78	&	1.04	&	0.44	&	0.24	\\
AGB 	&	697	&	 05140701-4003449 	&	--	&	14.73	&	1.02	&	0.17	&	-0.04	\\
AGB 	&	741	&	 05140758-4003164 	&	--	&	15.50	&	0.83	&	0.22	&	0.11	\\
AGB 	&	848	&	 05140883-4002380 	&	--	&	14.23	&	1.15	&	0.23	&	-0.05	\\
AGB 	&	849	&	 05140900-4004539 	&	--	&	14.83	&	0.97	&	0.41	&	0.21	\\
AGB 	&	887	&	 05140916-4002296 	&	--	&	14.99	&	0.87	&	0.27	&	0.09	\\
AGB 	&	988	&	 05141034-4004235 	&	--	&	14.76	&	0.98	&	0.70	&	0.49	\\
AGB 	&	989	&	 05141026-4003150 	&	--	&	15.42	&	0.80	&	0.17	&	0.05	\\
AGB 	&	1004	&	 05141057-4003308 	&	--	&	14.93	&	0.94	&	0.33	&	0.15	\\
AGB 	&	1014	&	 05141035-3958148 	&	--	&	14.72	&	1.04	&	0.57	&	0.36	\\
AGB 	&	1037	&	 05141084-4001475 	&	--	&	15.19	&	0.88	&	0.52	&	0.37	\\
AGB 	&	1172	&	 05141351-4003408 	&	--	&	14.87	&	0.96	&	0.66	&	0.47	\\
AGB 	&	1214	&	 05141501-4003040 	&	--	&	14.85	&	0.97	&	0.39	&	0.20	\\
AGB 	&	1246	&	 05141641-4002214 	&	--	&	15.42	&	0.80	&	0.29	&	0.17	\\
\tableline\tableline
    \end{tabular}}
\end{center}
\end{table*}

In the upper panel of Figure \ref{fig:doubleplot} we show a kernel density
estimate (KDE) histogram of $\delta$CN based on a Gaussian kernel with a
bandwidth of 0.035. This bandwidth was found to be optimal taking into
account the error bars and small number statistics. It was tested with many
other similar datasets (Campbell et al., in prep.). It can be seen in
Figure \ref{fig:doubleplot} that the distribution of $\delta$CN in the RGB
stars is \emph{quadrimodal}, having four peaks. Changing the KDE bandwidth
within reasonable limits ($\sim 0.02 \rightarrow 0.05$) does not alter this
result. This was an unexpected result because most clusters in our greater
sample and in the literature show bimodal distributions. As a comparison,
in the right-hand panels of Figure \ref{fig:doubleplot} we show the same
plot but for NGC 288. The data for this cluster were taken in the same
observing run and with the same instrument as our NGC 1851 data. We chose
this cluster for comparison because it has a similar metallicity to NGC
1851 ([Fe/H$]\sim -1.3$ versus $-1.2$, respectively) but a different
horizontal branch (HB) morphology -- NGC 1851 has a bimodal HB morphology
while NGC 288 has a blue HB only. As can be seen in this Figure we find the
standard bimodality in $\delta$CN in the RGB population of NGC 288.

There is the possibility that the observed quadrimodality in our sample of
NGC 1851 RGB stars is a chance occurrence due to a small sample size. A
small random sample drawn from a bimodal distribution may give this
result. To estimate the likelihood of this happening we conducted the
following test. We established an arbitrary $\delta$CN distribution
consisting of two Gaussians with centres at 0.05 and 0.45 and FWHMs of
0.10. Samples of 17 stars were then randomly drawn from this distribution
and the pseudo data was then smoothed with a Gaussian kernel having a
bandwidth of 0.035. We repeated this 1000 times. The results from this test
indicate that a small fraction ($\sim 2.5\%$) of the randomly generated
samples do present quadrimodal distributions (although only 1 out of 1000
was as clearly defined as in the real data). Therefore we cannot eliminate
the small probability that the observed $\delta$CN distribution in the RGB
population has been drawn from a bimodal population through chance. However
the case for quadrimodality is strengthened by the AGB results in Figure
\ref{fig:doubleplot} -- the AGB population \emph{also} appears to be
quadrimodal in $\delta$CN. Here the the stars are distributed differently
though, with the majority of the AGB stars being in the two CN-weaker
subpopulations, as is usually the case in GCs (e.g. NGC 288 in
Fig. \ref{fig:doubleplot}; \citealt{NCF81,paper3}). The probability for
attaining a quadrimodal distribution in the RGB population \emph{and
  simultaneously} in the AGB population by chance then becomes extremely
small, since they are essentially independent populations with different
internal distributions of $\delta$CN.

It has been suggested that NGC 1851 may be a merger product between two
GCs, initially as an explanation for producing its bimodal HB
\citep{Vandenbergh96,Catelan97}. In this scenario it would also be expected
that each merging population would have two `normal' subpopulations, each
with its own C-N and O-Na (and possibly Mg-Al) anticorrelations, and that
the superimposition of these populations would present dual
anticorrelations.  In the RGB study by \cite{Carretta11} it is
indeed found that NGC 1851 has two Na-O anticorrelations, one in their
metal-rich population and one in their metal-poor population.  This ties in
well with the findings of the \cite{Gratton12HB} study where it was found
that there are also two independent Na-O anticorrelations on the horizontal
branch -- one in the RHB population and one in the BHB population.  With
regards to C and N the picture is less clear. \cite{Lardo12} studied the
SGB populations and found a spread in C and N between stars. These elements
were also found to be anticorrelated, however no bimodal signature was
detected. The resolution of the spectra in that study was however quite
low, with R $\sim 1000$.  In the SGB study of \cite{Gratton12SGB} it was
found that the two subgiant branches have different average C
abundances. Interestingly they also found that there are different
proportions of C-normal and C-poor stars in each SGB, which may indicate
that each SGB hosts multiple subpopulations, again suggestive of a merger
scenario. In a study of the two RGB populations \cite{Villanova10} found a
spread in CNO elements but found no difference in C+N+O between the
populations.  \cite{Yong09} did however find a significant variation (a
factor of 4) in C+N+O in four RGB stars. Thus there is still uncertainty as
to whether C+N+O is constant between populations or not. This is an
important diagnostic since it is a very useful discriminant between
possible polluters of the primordial material from which N-rich populations
form. Cluster age determinations are also very sensitive to C+N+O
\citep{Rood85,Cassisi08}. More information/observations of CNO are needed
to clarify the situation.

In the case of CN bandstrengths the merger scenario leads to a natural
expectation that the two bimodal CN populations in the original clusters
would also superimpose, giving a quadrimodal distribution. Thus our
discovery of a quadrimodal distribution of CN band strengths in the RGB and
AGB populations of NGC 1851 adds further weight to the merger formation
scenario for this cluster. The CN quadrimodality also suggests that there
may be four populations with different N abundances in this
cluster. Although CN is generally accepted as a proxy for N we note that
the band strengths may also be affected by the abundances of C and O. Thus,
again, a complete set of (absolute) abundance observations including C, N,
and O are needed.


\begin{figure}[!th]
\centering
\includegraphics[width=0.5\columnwidth]{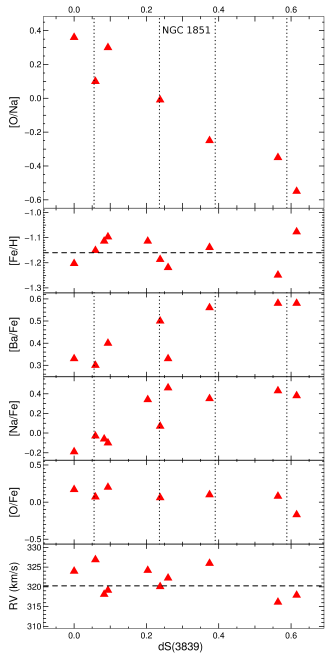}
\caption{\footnotesize{NGC 1851 RGB stars in common with the
    \cite{Carretta11} catalogue. Various abundance ratios plus radial
    velocity from the catalogue are shown versus our $\delta$S(3839)
    values. The vertical dotted lines denote the peaks of the 4 $\delta$CN
    populations (see Fig. \ref{fig:doubleplot}). The horizontal lines in
    [Fe/H] and radial velocity (RV) show the cluster means of $-1.16$ and
    $+320.26$ km/s respectively \citep{Carretta11}.}}
\label{fig:xmatch}
\end{figure}

With a view to gaining more information about the four $\delta$CN
populations we performed a cross-match of our sample with the
\cite{Carretta11} catalogue of RGB star abundances. The catalogue contains
124 stars with a range of abundance measurements for p-capture,
$\alpha$-capture, Fe-peak, and n-capture elements. Our cross-match found an
overlap of 10 RGB stars and one AGB star (Table
\ref{tab:observations}). Not all of the \cite{Carretta11} objects have the
full set of abundance results but for some abundances we have 7-10 stars to
compare with. In Figure \ref{fig:xmatch} we show various abundance ratios
from \cite{Carretta11} against our RGB $\delta$CN values. In the second
panel of Figure \ref{fig:xmatch} it can be seen that there appears to be no
correlation between our four $\delta$CN groups and [Fe/H]. This is contrary
to what we might expect from the SGB results of \cite{Gratton12SGB} where
it was found that the two SGB populations differ in average [Fe/H]. We note
however that the cross-matched sample is very small, especially within each
of the four subpopulations, which contain only 1-4 stars each. Sodium on
the other hand shows a definite correlation with $\delta$CN. This is
typical of GC abundance anomalies, where N is higher in the stars with high
Na. Barium also appears to correlate with $\delta$CN. In contrast oxygen
shows little variation with $\delta$CN. By using the ratio [O/Na] the
`noise' of the Fe scatter can be removed, and the relationship between O
and Na is amplified. We show [O/Na] in the top panel of Figure
\ref{fig:xmatch}. Here there is a striking anticorrelation, such that the
CN-weak populations show much higher O/Na ratios than the CN-strong
populations.  The lack of correlation between Fe (and Ca, Si, Ti - not
shown) and the light elements + s-process elements suggests that the
Fe-group (+ $\alpha$ capture) nucleosynthetic source(s) are separate to the
light element + s-process source(s). AGB stars are the most likely
primordial source for the O, Na, CN and s-process enhancements. If the four
populations are indeed real, then this suggests that each population was
polluted by AGB star ejecta to differing degrees.  We note that the star
with intermediate [Ba/Fe$] \sim 0.5$ is the only star in the sample with no
K magnitude from the \cite{Carretta11} database and thus may be an
unreliable data point. If so, then Ba would present a \emph{bimodal}
distribution (panel 3 of Fig. \ref{fig:xmatch}). Importantly each mode of
Ba abundance would be associated with two $\delta$CN peaks: the two
CN-weaker populations would be Ba-poor compared to the two
$\delta$CN-richer populations. This could be a useful diagnostic for
disentangling the multiple populations in the merger scenario. Clear
bimodalities in s-process abundances have been reported for the RGB
(\citealt{Yong08,Villanova10}) and the SGBs
(\citealt{Gratton12SGB}). Interestingly \cite{Carretta11} do not report a
bimodality in Ba but do show that s-process elements are (anti)correlated
with p-capture elements.  In the bottom panel of Figure \ref{fig:xmatch} we
show the radial velocities for the cross-matched sample. It can be seen
that only the most extreme $\delta$CN population stands out, having an
average radial velocity $\sim 5$ km/s lower than the other three
populations, which might suggest this group is kinematically
distinct. Again we stress that this is a very small dataset so the
discussion above is only speculative. We note that our group is in the
process of collecting medium resolution, broad wavelength coverage spectra
using 2dF/AAomega to complement the excellent \cite{Carretta11} RGB dataset
with C and N abundances. When complete the combined dataset will allow a
`holistic' analysis (including absolute abundances of C, N, O and therefore
the sum C+N+O) of the abundance and population trends for NGC 1851 red
giants.

\section{Summary and Conclusions}
We have recorded a homogeneous set of spectra for 17 RGB and 21 AGB stars
in the globular cluster NGC 1851.  We find that the CN band strengths
divide into four groups in both the RGB and AGB populations. This lends
support to the theory that NGC 1851 formed from a merger of two clusters
since one of the expected signatures of this would be two superimposed
bimodal distributions in CN. 

We cross-matched our sample with that of the high resolution study of
\cite{Carretta11} and found a small number of stars in common. This gave us
the opportunity to compare elements that typically (anti)correlate with N
in globular clusters, such as O, Na and Ba. We found that Na did indeed
correlate with $\delta$CN. An anticorrelation [O/Fe] was less clear but
when considering [O/Na] it was found that there was a very strong
anticorrelation. A possible correlation with Ba was observed.  The
(anti)correlations between these elements and $\delta$CN (and thus
presumably N and hence C) suggest that the material from which each of the
four populations formed was polluted by AGB stars. We also speculated that
the Ba distribution may be bimodal, as found in previous studies. If so,
then the two CN-weaker and two CN-richer populations would be paired, and
this may reflect a distinction between the two GCs in the merger
hypothesis.  It must be noted that the comparison sample is small, so
strong conclusions could not be made. Large-sample, high-resolution
observations combining absolute abundances of C, N, O, Fe, neutron-capture
elements and radial velocities in the giant branches of NGC 1851 are needed
to check that there are indeed four chemically (or even kinematically)
distinct populations and to determine which subpopulations are related to
each other. Our group is in the process of collecting observational data to
this end.

Finally we note that the AGB samples we presented here are the largest AGB
samples in any GCs to date. Our finding that NGC 1851 and NGC 288 both have
CN-weak dominated AGB populations adds to a growing picture in the
literature that the AGB CN distributions in GCs are different to the RGB
distributions \citep{paper2,paper3,Lai11,SMB11,Simpson12}.  For brief
reviews on this topic see \cite{SIK00} and \cite{paper1}. The current study
forms part of a larger study with relatively large datasets of AGB stars to
confirm this for a range of GCs (Campbell et al., in prep.).


\acknowledgments SWC acknowledges support from the Australian Research
Council's Discovery Projects funding scheme (project DP1095368).  {\it
  Facilities:} \facility{AAT}.

\bibliographystyle{apj}


\end{document}